\documentclass{elsart}
\usepackage{graphicx,amssymb,amsmath}
\usepackage[a4paper]{geometry}
\usepackage{ulem}
\usepackage{color}
\usepackage{rotating}
\journal{Nuclear Physics B}

\begin{document}

\begin{frontmatter}

\title{Anisotropy of the interface tension of the three-dimensional Ising model} 
\author{E. Bittner, A. Nu{\ss}baumer \and W. Janke} 
\address{Institut f\"ur Theoretische Physik and Centre for
Theoretical Sciences (NTZ) -- Universit\"at Leipzig, 
Postfach 100\ 920, D-04009 Leipzig, Germany}

%  \shortauthor{E. Bittner \etal}
 
%\pacs{68.35.Rh}{Phase transitions and critical phenomena}
%\pacs{02.70.Uu}{Applications of Monte Carlo methods}
%\pacs{75.10.Hk}{Classical spin models} 
%      

\begin{abstract}
We determine the interface tension for the $100$, $110$ and $111$ interface 
of the simple cubic Ising model with nearest-neighbour interaction using novel
simulation methods. To overcome the droplet/strip transition and the droplet
nucleation barrier we use a newly developed combination of the multimagnetic 
algorithm with the parallel tempering method.
We investigate a large range of inverse temperatures to
study the anisotropy of the interface tension in detail.
\end{abstract}


\end{frontmatter}
%\maketitle
      
%%%%%%%%%%%%%%%%%%%%%%%%%%%%%%%%%%%%%%%%%%%%%%%%%%%%%%%%%%%%
\section{Introduction}

In many physical systems with discrete symmetry the anisotropy of the
interface tension can play an important role for various phenomena,
including equilibrium droplet shapes~\cite{droplet}
and the interfacial roughening transition~\cite{rough}.
For sufficiently strong anisotropy, facets, edges, or even corners can be
identified in the equilibrium droplet shape.  
Due to the anisotropy of the interface tension in the three-dimensional (3D) Ising model,
the shape of the equilibrium
droplet at some finite temperature is not spherical and has,
in principle, to be determined by the Wulff construction~\cite{wulff}. 
Since the 3D Ising model with nearest-neighbour interaction 
is not exactly solvable, no analytical results are available for the 
interfacial free energy and the Wulff construction can only be done using an
effective model of the angle-depending interface tension. Only for
temperatures not too far below the critical temperature one can use the 
spherical approximation and, therefore, it is important to know how large the
anisotropy is for a given temperature.

Whereas a lot of numerical results are available for the planar $100$ interface tension of the
simple cubic Ising model, see e.g. Refs.~\cite{3d_s100_binder,3d_s100_berg,hasenbusch94,3d_s100_caselle94,hasenbusch97,3d_s100_chatelain,3d_s100_caselle06},
there are only a few results in the literature for the $110$ interface~\cite{3d_s110} and, 
to our knowledge, there are no results at all for the $111$ interface. 

The layout of the remainder of this paper is organized as follows. In 
Sec.~\ref{planar} we discuss the results for the planar $100$ interface and
compare our results with previous estimates in the literature.
Next we describe in Sec.~\ref{tilt} first the special boundary conditions
employed for the simulation of the tilted interfaces with $110$ and $111$
orientation and then discuss the results of our finite-size scaling analysis.
Finally, in Sec.~\ref{summary} we conclude with a summary of our main findings.

%%%%%%%%%%%%%%%%%%%%%%%%%%%%%%%%%%%%%%%%%%%%%%%%%%%%%%%%%%%%
\section{Planar interface\label{planar}}
\begin{figure}
 \begin{center}
    \includegraphics[width=0.65\textwidth]{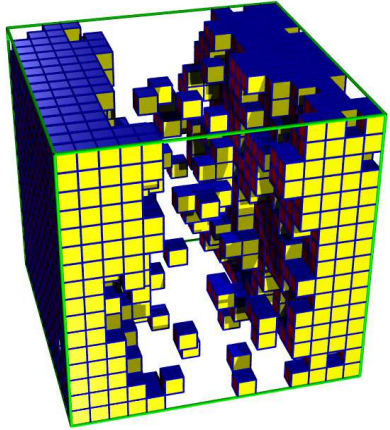}
 \end{center}
  \caption{Plot of a typical configuration with two $100$ interfaces ($\beta=0.3$).
  \label{fig:s100}
}
\end{figure}

We considered the Ising model on $L\times L \times L$ simple cubic lattices with periodic boundary conditions 
in all three directions to simulate systems with planar $100$ interfaces, 
for various temperatures below the
Ising transition at $\beta_c\equiv 1/T_c=0.22165459$~\cite{bloete99}.
%0.221\,654$~\cite{bloete95,talapov96,ballesteros99,hasenbusch99}. 
For a typical configuration see Fig.~\ref{fig:s100}. 

The interface tension can be measured using a multimagnetical
(flat in the distribution of the magnetization $m$)
simulation combined with parallel tempering~\cite{ebanwj},
the result of which is after appropriate reweighting to the canonical ensemble a 
double-peaked magnetization density $P(m)$. 
We simulated $n=26$ replica of the system at different inverse temperatures $\beta_i$, 
with $\beta_i=0.195+0.005 (i-1)$ and $i=1,\dots,n$. 
A planar interface of the 3D Ising model exhibits a transition at the 
roughening temperature~\cite{rough} $T_R=1/\beta_R$, with $\beta_R=0.407 58(1)$, above which the surface stiffness
for the $100$ interface is finite and below which it is infinite. 
Therefore, we restrict ourselves  to the temperature range above this
transition, i.e. $T_i=1/\beta_i>T_R$ for all $i$ .
To construct the weight function for the multimagnetic part of the algorithm,
we employed an accumulative recursion, described in detail in Refs.~\cite{wj_nic}
and \cite{berg96}. Statistical averages were taken over runs of
$1\times 10^6$ Monte Carlo (MC) steps, where one MC step 
consists of one full multimagnetical lattice sweep for all 26 replica and
one attempted parallel tempering exchange of all adjacent replica.

The interface tension $\sigma_{100}$ can be estimated according to
Refs.~\cite{binder81,3d_s100_binder} by ($d=3$)
\begin{equation}\label{eq:fs}
 \left(\frac{P_\mathrm{max}^{(L)}}{P_\mathrm{min}^{(L)}} \right)
 = A L^x \exp[2 L^{(d-1)} \beta \sigma_{100}] \;
\end{equation}
where $P_\mathrm{min}^{(L)}$ is the value of the magnetization density in the mixed phase
region $m \approx 0$ (strip phase) and $P_\mathrm{max}^{(L)}$ the value at its maxima located 
close to the equilibrium magnetization $m =\pm m_0$. 
Therefore, the finite-volume estimator is given by
\begin{equation}
\sigma_{100}(L)=
\frac{1}{2L^{d-1}\beta}
\ln\left(
\frac{P_\mathrm{max}^{(L)}}{P_\mathrm{min}^{(L)}}\right) =
\sigma_{100} 
+\frac{x \ln(L)}{2 L^{d-1} \beta}
+\frac{c_1}{L^{d-1}} 
\label{fit_sigma}
\end{equation}
with $c_1=(\ln A)/2\beta$.

The power of $L$ in the prefactor of Eq.~(\ref{eq:fs}) is a
delicate problem and the knowledge of the pre-exponential behaviour fixes one
free parameter of the fit. Using the capillary wave
approximation~\cite{brezin85,gelfand90,morris91} the exponent $x=(d-3)/2$,
i.e. $x=0$ for $d=3$. Therefore, we performed finite-size scaling fits according to 
\begin{equation}
\sigma_{100}(L) = \sigma_{100} 
+\frac{c_1}{L^{d-1}} \; ,
\label{fit_2}
\end{equation}
and, to allow for higher-order corrections, also to 
\begin{equation}
\sigma_{100}(L) = \sigma_{100} 
+\frac{c_1}{L^{d-1}} 
+\frac{c_2}{L^{2(d-1)}} \; .
\label{fit_3}
\end{equation}

\begin{figure}
 \begin{center}
    \includegraphics{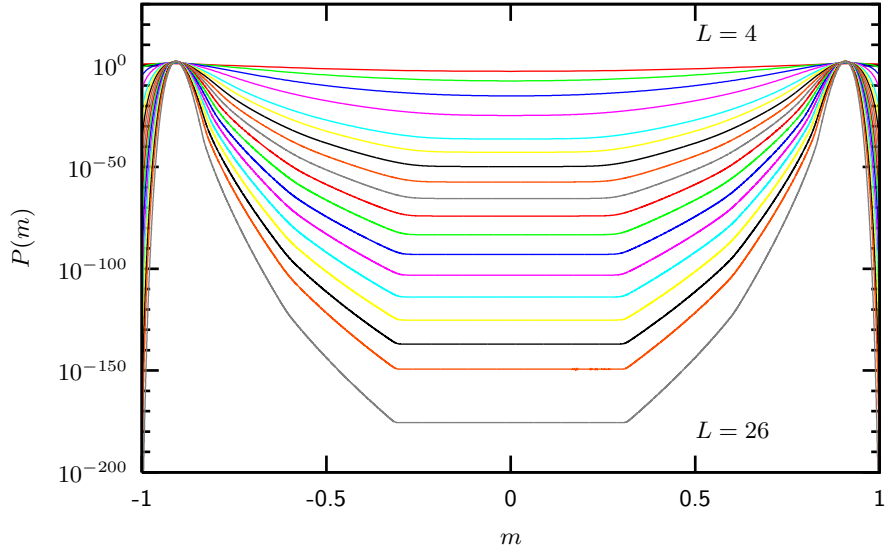}
 \end{center}
  \caption{Distribution of the magnetization $m$ for the 3D
Ising model with  periodic boundary conditions at $\beta=0.3$ and system sizes $L=4, \dots, 26$.
  \label{fig:p_m_s100}
}
\end{figure}

We performed simulations for various lattice sizes ranging 
from $L=4$ to $L=26$. In Fig.~\ref{fig:p_m_s100} we show the magnetization
density $P(m)$ for $\beta=0.3$, where the strip configurations, corresponding to
the minimum between the two peaks, are suppressed by more than 175 orders of
magnitude for the largest system and this suppression becomes
even more pronounced for lower temperatures. Such unlikely configurations 
would not be a problem for a multimagnetical algorithm, but
between the strip configuration and the droplet configuration there is an
exponentially large barrier~\cite{leung:90,neuhaus.hager} 
that might not be overcome during the
equilibration phase. Therefore, it is
necessary to use the combined algorithm to overcome this barrier. A similar
reasoning applies to the evaporation/condensation transition which is another 
hidden albeit weaker barrier in the multimagnetical
system~\cite{neuhaus.hager,binder03,nussbaum}.

\begin{figure}
 \begin{center}
    \includegraphics{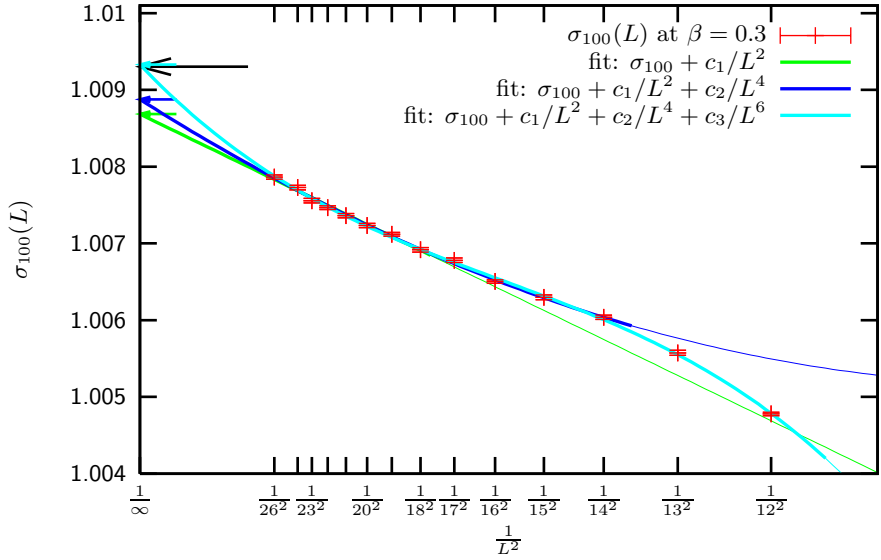}
 \end{center}
  \caption{Scaling of the interface-tension estimates from the histogram
  method for the inverse temperature $\beta=0.3$ and system sizes from $L=12$ up to $26$.
  The lines show the fits according to Eqs.~(\ref{fit_2}), (\ref{fit_3}),
and (\ref{fit_5}).
  The long black arrow on the $y$ axis points to the result of Hasenbusch and
Pinn~\cite{hasenbusch94} and
  the three short arrows indicate our fit results of $\sigma_{100}$. The thick lines indicate the fit range.
  \label{fig:fss_s100}
}
\end{figure}

\begin{figure}
\centerline{
\includegraphics[width=0.5\textwidth]{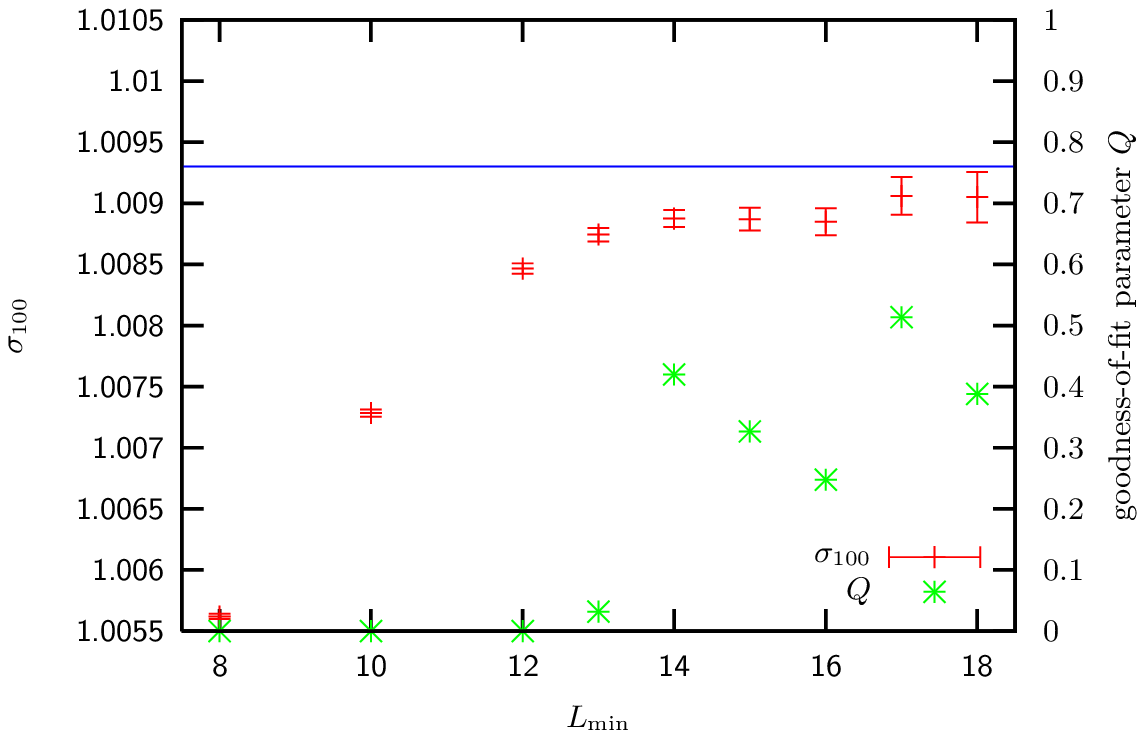}
\includegraphics[width=0.5\textwidth]{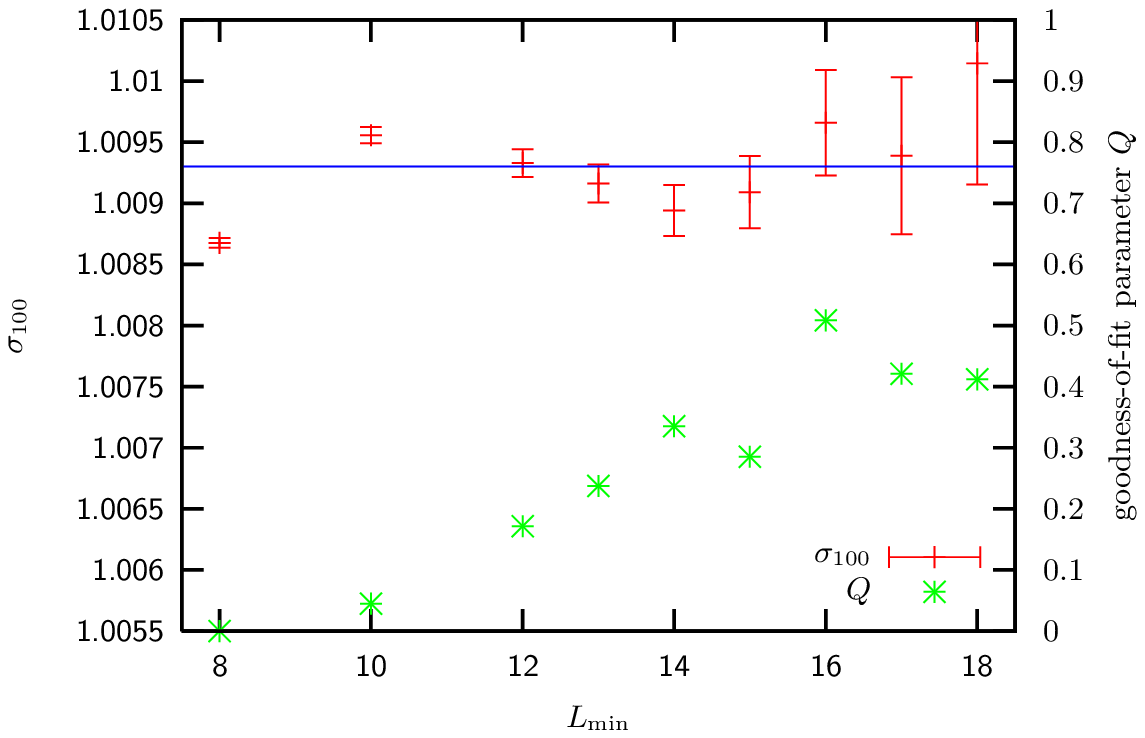}
}
  \caption{
  The interface tension $\sigma_{100}$ at $\beta=0.3$, as a function of the lower bounds $L_{\rm min}$ of the fit range,
  determined using fit ansatz Eq.~(\ref{fit_3}) (left) and Eq.~(\ref{fit_5}) (right), 
  respectively. 
  To judge the quality of the individual data points we also plot the goodness-of-fit parameter $Q$.
  \label{fig:sigma_s100}
}
\end{figure}

For every system the maximum and minimum probability
$P_\mathrm{max}^{(L)}$ and $P_\mathrm{min}^{(L)}$ were read off, and by
repeating the simulations $32$ times the statistical error bars were obtained.
For $\beta=0.3$ and $L \ge 12$ the resulting values for $\sigma_{100}(L)$ are plotted in
Fig.~\ref{fig:fss_s100}. 
To check the stability of the fit results we performed fits with different lower bounds $L_{\rm min}$
of the fit range. The upper bound of the fits was always 
the largest lattice $L=26$. For the fits according to Eq.~(\ref{fit_3}) we find due to the systematic variation of the
lower bound a trend to larger values of $\sigma_{100}$ with increasing $L_{\rm min}$. 
This can be seen in the left panel of Fig.~\ref{fig:sigma_s100} 
where we also include the goodness-of-fit parameter $Q$ into the figure to judge the quality of the fits.
Above $L_{\rm min}=14$ the goodness-of-fit parameter was well above $0.05$, which we chose as cutoff value. 
Nevertheless, not yet reaching a constant value for $\sigma_{100}$ led us to include one more parameter in our 
fits,
\begin{equation}
\sigma_{100}(L)=\sigma_{100}
+\frac{c_1}{L^{d-1}}
+\frac{c_2}{L^{2(d-1)}}+\frac{c_3}{L^{3(d-1)}}.
\label{fit_5}
\end{equation}
Performing fits with this ansatz and again varying the lower fit bound systematically, the resulting values for 
the planar interface tension $\sigma_{100}$ stay almost constant for reasonable fits ($Q\ge0.05$),
as one can see in the right panel of  Fig.~\ref{fig:sigma_s100}.
The infinite system size extrapolation in $1/L^2$ 
according to Eq.~(\ref{fit_5}) with $L_{\rm min}=12$ yields for the particular inverse temperature $\beta=0.3$ a value of
$\sigma_{100}=1.00933(12)$ 
for the planar interface tension with goodness-of-fit parameter $Q=0.18$, 
which is in good agreement with the result from Hasenbusch
and Pinn~\cite{hasenbusch94} $\sigma_{100}=1.009302(106)$.
All results with $Q\ge0.05$ for the three different infinite system size extrapolations of the planar interface tension are collected
in Tables~\ref{tab:s100} and ~\ref{tab:s100ci}. 

\begin{sidewaystable}
  \caption{
Results for the planar interface tension $\sigma_{100}$ using fits according to
Eqs.~(\ref{fit_2}), (\ref{fit_3}), and (\ref{fit_5}), respectively. 
In the last column we include for comparison the results of Hasenbusch  
and Pinn~\cite{hasenbusch94}.
}
  \label{tab:s100}
 \begin{tabular}{lllllllrlll} 

  \hline\hline

  &\multicolumn{3}{c}{fit ansatz (\ref{fit_2})}       
  &\multicolumn{3}{c}{fit ansatz (\ref{fit_3})}
  &\multicolumn{3}{c}{fit ansatz (\ref{fit_5})}
  &\multicolumn{1}{c}{Ref.~\cite{hasenbusch94}}\\       

   \multicolumn{1}{c}{$\beta$}       
  &\multicolumn{1}{c}{fit range}       
  &\multicolumn{1}{c}{$Q$}       
  &\multicolumn{1}{c}{$\sigma_{100}$}       
  &\multicolumn{1}{c}{fit range}       
  &\multicolumn{1}{c}{$Q$}       
  &\multicolumn{1}{c}{$\sigma_{100}$}       
  &\multicolumn{1}{c}{fit range}       
  &\multicolumn{1}{c}{$Q$}       
  &\multicolumn{1}{c}{$\sigma_{100}$}       
  &\multicolumn{1}{c}{$\sigma_{100}$}  \\ 

  \hline
0.265  & $19-26$ & 0.32 & 0.60073(5) & $17-26$ &  0.10 & 0.60015(15) & $14-26$ & 0.34 &  0.60175(21) &0.601124(194)\\
0.27   & $17-26$ & 0.47 & 0.66586(4) & $17-26$ &  0.38 & 0.66579(15) & $14-26$ & 0.13 &  0.66679(21) &0.666354(180)\\
0.275  & $17-26$ & 0.36 & 0.72862(4) & $17-26$ &  0.55 & 0.72885(15) & $12-26$ & 0.15 &  0.72971(11) &0.729214(146)\\
0.28   & $18-26$ & 0.52 & 0.78920(5) & $17-26$ &  0.87 & 0.78962(15) & $13-26$ & 0.13 &  0.79029(11) &0.789788(142)\\
0.285  & $18-26$ & 0.34 & 0.84742(5) & $17-26$ &  0.84 & 0.84787(15) & $10-26$ & 0.14 &  0.84823(7)  &0.848066(132)\\
0.29   & $14-26$ & 0.36 & 0.90330(3) & $14-26$ &  0.46 & 0.90339(7)  & $12-26$ & 0.22 &  0.90439(7)  &0.903996(122)\\
0.295  & $15-26$ & 0.17 & 0.95702(3) & $14-26$ &  0.78 & 0.95723(8)  & $10-26$ & 0.11 &  0.95808(8)  &0.957756(126)\\
0.3    & $16-26$ & 0.09 & 1.00861(3) & $14-26$ &  0.41 & 1.00888(7)  & $12-26$ & 0.18 &  1.00933(12) &1.009302(106)\\
0.305  & $16-26$ & 0.14 & 1.05808(4) & $13-26$ &  0.09 & 1.05826(6)  & $10-26$ & 0.10 &  1.05894(7)  &1.058752(94)\\ 
0.31   & $16-26$ & 0.25 & 1.10550(3) & $13-26$ &  0.27 & 1.10574(6)  & $8 -26$ & 0.14 &  1.10618(5)  &1.106152(88)\\ 
0.315  & $16-26$ & 0.24 & 1.15095(3) & $13-26$ &  0.08 & 1.15121(6)  & $8 -26$ & 0.05 &  1.15168(5)  &1.151608(76)\\ 
0.32   & $16-26$ & 0.05 & 1.19446(4) & $13-26$ &  0.08 & 1.19474(6)  & $10-26$ & 0.12 &  1.19508(8)  &1.195140(62)\\ 

  \hline\hline
 \end{tabular}
%}
\end{sidewaystable}

\begin{sidewaystable}
  \caption{Results for the parameters $c_i$ for the planar interface tension $\sigma_{100}$ 
from fits according to Eqs.~(\ref{fit_2}), (\ref{fit_3}), and (\ref{fit_5}), respectively.
}
  \label{tab:s100ci}
 \begin{tabular}{llllllllrllll}

  \hline\hline

  &\multicolumn{3}{c}{fit ansatz (\ref{fit_2})}
  &\multicolumn{4}{c}{fit ansatz (\ref{fit_3})}
  &\multicolumn{5}{c}{fit ansatz (\ref{fit_5})}\\

   \multicolumn{1}{c}{$\beta$}
  &\multicolumn{1}{c}{fit range}
  &\multicolumn{1}{c}{$Q$}
  &\multicolumn{1}{c}{$c_1$}
  &\multicolumn{1}{c}{fit range}
  &\multicolumn{1}{c}{$Q$}
  &\multicolumn{1}{c}{$c_1$}
  &\multicolumn{1}{c}{$c_2$}
  &\multicolumn{1}{c}{fit range}
  &\multicolumn{1}{c}{$Q$}
  &\multicolumn{1}{c}{$c_1$}
  &\multicolumn{1}{c}{$c_2$}  
  &\multicolumn{1}{c}{$c_3$}  \\

  \hline
0.265  & $19-26$ & 0.32 & $-0.70(3)$ & $17-26$ &  0.10 & $-0.13(13)$ & $-139(26)$ & $14-26$ & 0.34 & $-2.3(2)  $  & 812(69) & $-132216(6968)$ \\
0.27   & $17-26$ & 0.47 & $-0.67(2)$ & $17-26$ &  0.38 & $-0.61(13)$ & $-13(26) $ & $14-26$ & 0.13 & $-2.0(2)  $  & 615(72) & $-89832(7232) $ \\
0.275  & $17-26$ & 0.36 & $-0.60(2)$ & $17-26$ &  0.55 & $-0.81(13)$ & $41(25)  $ & $12-26$ & 0.15 & $-1.97(10)$  & 549(26) & $-71112(2067) $ \\
0.28   & $18-26$ & 0.52 & $-0.61(2)$ & $17-26$ &  0.87 & $-1.00(13)$ & $87(25)  $ & $13-26$ & 0.13 & $-1.72(15)$  & 420(43) & $-49729(3858) $ \\
0.285  & $18-26$ & 0.34 & $-0.60(2)$ & $17-26$ &  0.84 & $-1.02(13)$ & $94(25)  $ & $10-26$ & 0.14 & $-1.55(5) $  & 345(10) & $-37817(554)  $ \\
0.29   & $14-26$ & 0.36 & $-0.55(1)$ & $14-26$ &  0.46 & $-0.61(5) $ & $10(7)   $ & $12-26$ & 0.22 & $-1.46(10)$  & 291(26) & $-28835(2042) $ \\
0.295  & $15-26$ & 0.17 & $-0.53(1)$ & $14-26$ &  0.78 & $-0.69(5) $ & $27(7)   $ & $10-26$ & 0.11 & $-1.58(5) $  & 308(11) & $-27491(621)  $ \\
0.3    & $16-26$ & 0.09 & $-0.54(1)$ & $14-26$ &  0.41 & $-0.75(5) $ & $38(8)   $ & $12-26$ & 0.18 & $-1.24(10)$  & 202(26) & $-16903(2078) $ \\
0.305  & $16-26$ & 0.14 & $-0.54(1)$ & $13-26$ &  0.09 & $-0.69(4) $ & $29(5)   $ & $10-26$ & 0.10 & $-1.36(5) $  & 224(10) & $-17335(595)  $ \\ 
0.31   & $16-26$ & 0.25 & $-0.54(1)$ & $13-26$ &  0.27 & $-0.73(4) $ & $37(5)   $ & $8 -26$ & 0.14 & $-1.19(3) $  & 175(4)  & $-12841(155)  $ \\ 
0.315  & $16-26$ & 0.24 & $-0.55(1)$ & $13-26$ &  0.08 & $-0.76(4) $ & $41(5)   $ & $8 -26$ & 0.05 & $-1.22(3) $  & 174(4)  & $-11695(158)  $ \\ 
0.32   & $16-26$ & 0.05 & $-0.54(1)$ & $13-26$ &  0.08 & $-0.77(4) $ & $43(5)   $ & $10-26$ & 0.12 & $-1.10(5) $  & 141(11) & $-8826(620)   $ \\ 

  \hline\hline
 \end{tabular}
%}
\end{sidewaystable}

%%%%%%%%%%%%%%%%%%%%%%%%%%%%%%%%%%%%%%%%%%%%%%%%%%%%%%%%%%%%
\section{Tilted interfaces\label{tilt}}
\begin{figure}
 \begin{center}
    \includegraphics[width=0.65\textwidth]{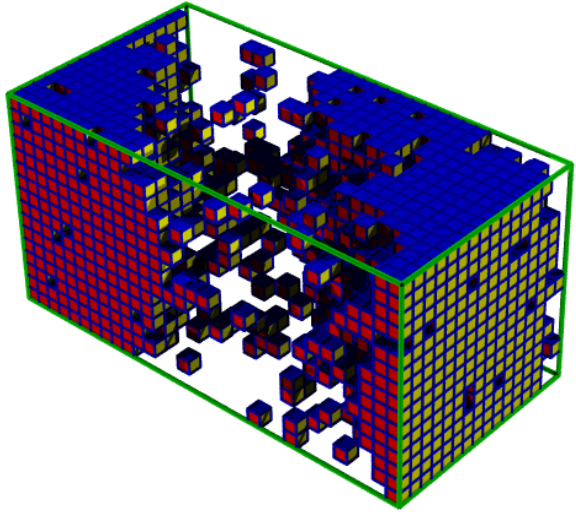}
 \end{center}
  \caption{Plot of a typical configuration with two $110$ interfaces
($\beta=0.3$).
  \label{fig:s110} 
} 
\end{figure}

To generate a tilted interface we used a $2L\times L\times L$ simple cubic lattice under two sets
of boundary conditions. We chose a rather unusual combination of boundary conditions for this
study. For the lattice with an interface along the $110$ direction we imposed  periodic
boundary conditions in the $x$ and $y$ direction and shifted boundary condition in the
$z$ direction. To be more precise, the neighbour in negative $z$ direction of a spin in the first $xy$
layer of the system $s(x,y,1)$ is $s(x+L,y,L)$ for $1\le x\le L$ and $s(x,y,L)$ for $L< x\le 2L$.
A typical configuration for such a system below the Ising transition is depicted in Fig.~\ref{fig:s110}.

\begin{figure}
 \begin{center}
    \includegraphics[width=0.65\textwidth]{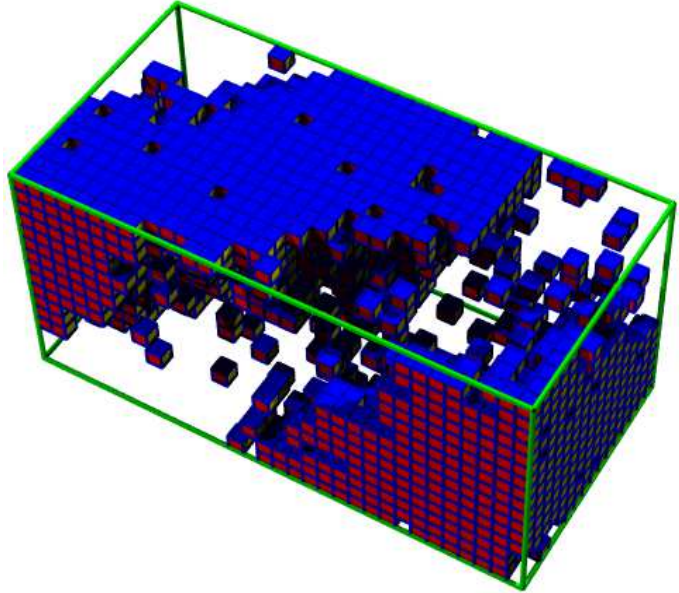}
 \end{center}
  \caption{Plot of a typical configuration with two $111$ interfaces
($\beta=0.3$).
  \label{fig:s111}
}
\end{figure}

For the lattice with an interface along the $111$ direction we imposed  periodic
boundary conditions in the $x$ direction and shifted boundary condition in the
$y$ and $z$ direction. Therefore, the neighbour in negative $y$ direction of a spin in the first $xz$
layer of the system $s(x,1,z)$ is $s(x+L,L,z)$ for $1\le x\le L$ and $s(x,L,z)$ for $L< x\le 2L$.
A typical configuration for such a system below the Ising transition is depicted in Fig.~\ref{fig:s111}.  

\begin{figure}
\includegraphics{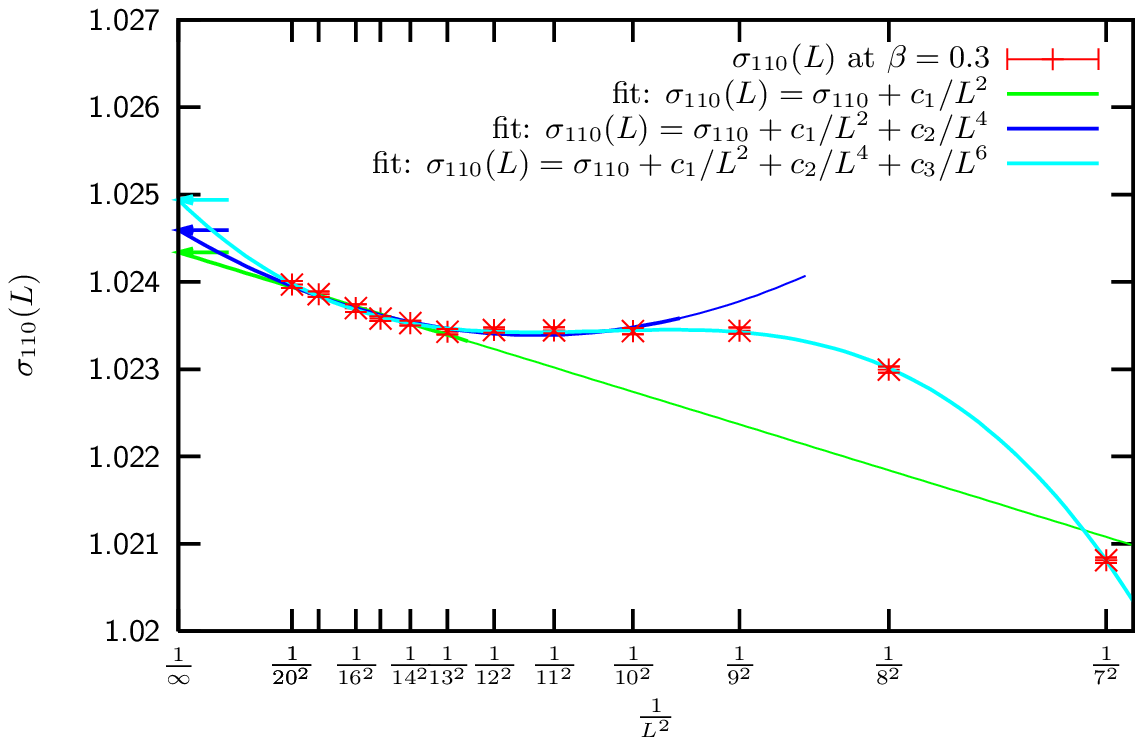}
  \caption{Scaling of estimates for the interface tension $\sigma_{110}$ from the histogram
  method for the inverse temperature $\beta=0.3$ and system sizes from $L=7$ up to $20$.
  The lines show the fits according to Eqs.~(\ref{fit_2}), (\ref{fit_3}), and (\ref{fit_5}).
  The small arrows on the $y$-axes indicate the fit results of $\sigma_{110}$ and 
  the thick lines indicate the fit range.
  \label{fig:fss_s110}
} 
\end{figure}

\begin{figure}
    \includegraphics{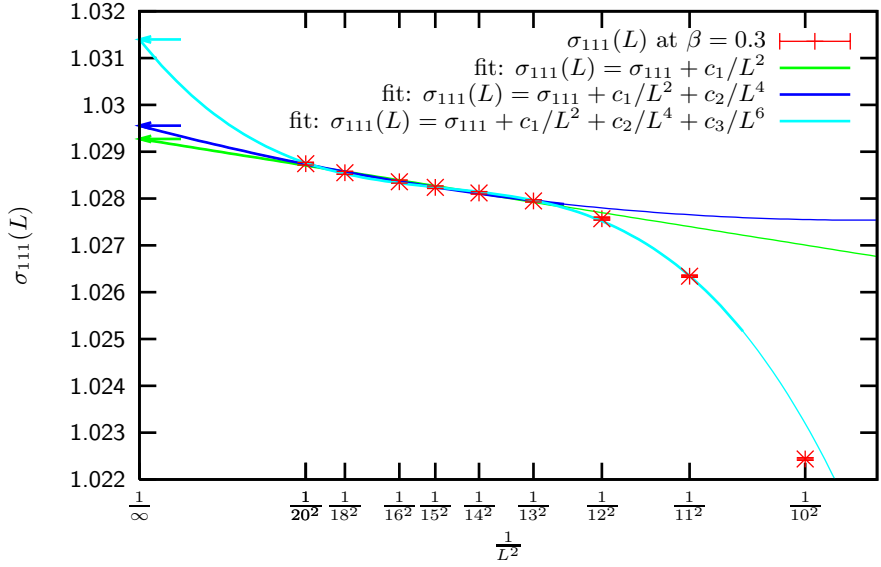}
  \caption{Same as Fig.~\ref{fig:fss_s110} for the interface tension
$\sigma_{111}$ at $\beta=0.3$ and system sizes from $L=10$ up to $20$.
  \label{fig:fss_s111}
}
\end{figure}

Using the same setup as described above we measured the probability density of the magnetization 
for various lattice sizes ranging from $L=4$ to $L=20$. From these distributions, we determined the
interface tension in the $110$ and $111$ direction by means of
infinite-system size extrapolations via Eqs.~(\ref{fit_2}), (\ref{fit_3}), and (\ref{fit_5}).
In Figs.~\ref{fig:fss_s110} and \ref{fig:fss_s111} we show the different
fits for $\beta=0.3$. 
To check the stability of the fit results we again performed fits with different lower bounds $L_{\rm min}$
of the fit range. The upper bound of the fits was always 
the largest lattice $L=20$. 
Similarly to $\sigma_{100}$, the fits according to Eq.~(\ref{fit_3}) show a slight trend to larger values
of $\sigma_{110}$ and $\sigma_{111}$ with increasing $L_{\rm min}$ whereas for the ansatz of Eq.~(\ref{fit_5}) the interface tensions
stay almost constant within error bars for reasonable fits ($Q>0.05$).
All results with $Q\ge0.05$ are collected in Tables~\ref{tab:s110}, \ref{tab:s110ci}, \ref{tab:s111}, and \ref{tab:s111ci},  respectively. 

For $\sigma_{111}$ the difference between the final estimates for the interface tension using the 
different infinite-volume extrapolations is larger than for the other two interface directions. 
This is an indication that the finite-size effects are more pronounced in this
setup. One reason for this is that the distance between the two interfaces is
too small and therefore the fluctuations of the interfaces are correlated and the
effect becomes more pronounced as the critical temperature is approached.
In spite of the computational effort, the system sizes are still too small to
give equally accurate values for $\sigma_{111}$ as for $\sigma_{100}$ and
$\sigma_{110}$, respectively. Nevertheless, the statistical error of the interface
tension of the 111 interface is only roughly $0.01\% - 0.04\%$, which is more than one order of
magnitude smaller than the effect which we are interested in, namely the anisotropy
of the interface tension.

\begin{figure}
    \includegraphics{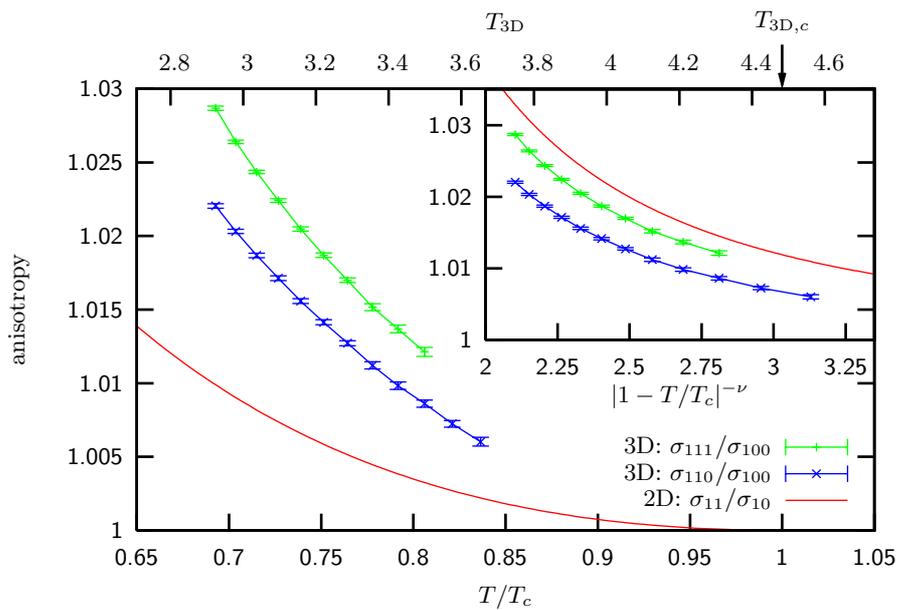}
  \caption{
The anisotropy of the interface tension for the $110$ and $111$ direction as a function of the 
temperature $T=1/\beta$ (upper scale), respectively. When approaching the 3D
Ising transition temperature indicated by the arrow, 
the interface tension becomes isotropic and  finally vanishes.   
To allow a comparison with the 2D Ising model, the lower scale shows the 
reduced temperature $T/T_c=\beta_c/\beta$. The inset shows another comparison,
where the abscissa is proportional to the asymptotic correlation length $\xi \propto |
1-T/T_c|^{-\nu}$ (with $\nu=1$ in 2D and $\nu=0.63$ in 3D).
  \label{fig:ratio_2d_3d}
}
\end{figure}

Using the results from the fits according to Eq.~(\ref{fit_2}) we determined the anisotropy 
of the interface tension for different directions, namely the $110$ and $111$ direction.
Due to the large finite-size effects near the critical point, the accessible temperature range is limited to 
a window at low temperatures. Therefore, we cannot give reliable results for
the region near the transition. In Fig.~\ref{fig:ratio_2d_3d} we show the anisotropy as  a function of the reduced temperature
and include for comparison results for the two-dimensional Ising model. 
In the two-dimensional case we calculated the anisotropy using the exact expressions for the
interface tension of the $10$ ``surface''~\cite{onsager},
\begin{equation}
 \sigma_{10} = 2 + \frac 1 \beta \ln [\tanh(\beta)],
 \label{eq:s10}
\end{equation}
and for the 11 ``surface''~\cite{baxter_book},
\begin{equation}
\sigma_{11} = \frac{\sqrt 2}\beta \ln \left[ \sinh \left( 2
\beta \right) \right].
 \label{eq:s11}
\end{equation}
Our results show that the anisotropy of the interface tension
as a function of the reduced temperature grows faster in three dimensions than
in two. Nevertheless, the absolute
value is still very small ($<3\%$) for the temperature range investigated in this work.

%%%%%%%%%%%%%%%%%%%%%%%%%%%%%%%%%%%%%%%%%%%%%%%%%%%%%%%%%%%%
\section{Summary\label{summary}}

We have presented a careful analysis of the interface tension in $100$, $110$
and $111$ direction of the simple cubic Ising model with nearest-neighbour
interaction. Using a newly developed combination of the multimagnetic
algorithm with the parallel tempering method, we were able to measure the
highly suppressed configurations of the strip phase for systems up to
$26^3$ at $\beta=0.32$, i.e. $T\sim 0.7~T_c$. 
We show that at given $T/T_c$ the anisotropy of the
interface in three dimensions is larger than in two dimensions.
However, down to $0.7~T_c$ it never exceeds $3\%$, so that in most cases the
isotropic approximation for droplet condensation phenomena should be
sufficiently accurate.

%%%%%%%%%%%%%%%%%%%%%%%%%%%%%%%%%%%%%%%%%%%%%%%%%%%%%%%%%%%%
\section{Acknowledgements}
We wish to thank Martin Hasenbusch for helpful discussions.

Work supported by the Deutsche Forschungsgemeinschaft (DFG) under grants No.\
JA483/22-1 and No.\ JA483/23-1 and in part by the EU RTN-Network `ENRAGE':
``Random Geometry and Random Matrices: From Quantum Gravity to Econophysics''
under grant No.~MRTN-CT-2004-005616. Supercomputer time at NIC J{\"u}lich under
grant No.~hlz10 is also gratefully acknowledged.

\newpage

\begin{sidewaystable}
   \caption{ Results for the $110$ interface tension using of fits
according to
Eqs.~(\ref{fit_2}), (\ref{fit_3}), and (\ref{fit_5}), respectively. 
}
 \label{tab:s110}
 \begin{tabular}{llllrlllll}

  \hline\hline

  &\multicolumn{3}{c}{fit ansatz (\ref{fit_2})}
  &\multicolumn{3}{c}{fit ansatz (\ref{fit_3})}
  &\multicolumn{3}{c}{fit ansatz (\ref{fit_5})}\\

   \multicolumn{1}{c}{$\beta$}
  &\multicolumn{1}{c}{fit range}
  &\multicolumn{1}{c}{$Q$}
  &\multicolumn{1}{c}{$\sigma_{110}$}
  &\multicolumn{1}{c}{fit range}
  &\multicolumn{1}{c}{$Q$}
  &\multicolumn{1}{c}{$\sigma_{110}$}
  &\multicolumn{1}{c}{fit range}
  &\multicolumn{1}{c}{$Q$}
  &\multicolumn{1}{c}{$\sigma_{110}$} \\

  \hline

0.265 & $12-20$ &0.19 & 0.60441(5) & $12-20$& 0.15&0.60430(14) & $9-20$ &0.32 & 0.60496(13)\\
0.27  & $11-20$ &0.08 & 0.67068(4) & $12-20$& 0.71&0.67093(14) & $9-20$ &0.96 & 0.67161(15)\\
0.275 & $13-20$ &0.22 & 0.73490(6) & $12-20$& 0.94&0.73532(14) & $9-20$ &0.88 & 0.73589(15)\\
0.28  & $13-20$ &0.23 & 0.79696(6) & $11-20$& 0.33&0.79722(11) & $8-20$ &0.41 & 0.79769(11)\\
0.285 & $13-20$ &0.16 & 0.85692(6) & $11-20$& 0.37&0.85724(11) & $8-20$ &0.93 & 0.85766(11)\\
0.29  & $13-20$ &0.06 & 0.91478(6) & $10-20$& 0.08&0.91500(8)  & $7-20$ &0.43 & 0.91539(8)\\
0.295 & $13-20$ &0.28 & 0.97055(6) & $10-20$& 0.25&0.97077(8)  & $7-20$ &0.83 & 0.97113(7)\\
0.3   & $13-20$ &0.26 & 1.02434(6) & $10-20$& 0.11&1.02459(8)  & $7-20$ &0.84 & 1.02494(8)\\
0.305 & $13-20$ &0.35 & 1.07621(6) & $10-20$& 0.14&1.07645(8)  & $7-20$ &0.60 & 1.07679(7)\\
0.310 & $13-20$ &0.05 & 1.12625(6) & $10-20$& 0.09&1.12645(8)  & $7-20$ &0.46 & 1.12668(8)\\
0.315 & $13-20$ &0.33 & 1.17433(6) & $9 -20$& 0.13&1.17449(7)  & $6-20$ &0.46 & 1.17474(6)\\
0.32  & $13-20$ &0.21 & 1.22079(6) & $10-20$& 0.31&1.22112(9)  & $6-20$ &0.39 & 1.22120(6)\\

  \hline\hline
 \end{tabular}

\end{sidewaystable}

\begin{sidewaystable}
  \caption{Results for the parameters $c_i$ for the $110$ interface tension 
from fits according to Eqs.~(\ref{fit_2}), (\ref{fit_3}), and (\ref{fit_5}), respectively.
}
  \label{tab:s110ci}
 \begin{tabular}{llllrlllrllll}

  \hline\hline

  &\multicolumn{3}{c}{fit ansatz (\ref{fit_2})}
  &\multicolumn{4}{c}{fit ansatz (\ref{fit_3})}
  &\multicolumn{5}{c}{fit ansatz (\ref{fit_5})}\\

   \multicolumn{1}{c}{$\beta$}
  &\multicolumn{1}{c}{fit range}
  &\multicolumn{1}{c}{$Q$}
  &\multicolumn{1}{c}{$c_1$}
  &\multicolumn{1}{c}{fit range}
  &\multicolumn{1}{c}{$Q$}
  &\multicolumn{1}{c}{$c_1$}
  &\multicolumn{1}{c}{$c_2$}
  &\multicolumn{1}{c}{fit range}
  &\multicolumn{1}{c}{$Q$}
  &\multicolumn{1}{c}{$c_1$}
  &\multicolumn{1}{c}{$c_2$}
  &\multicolumn{1}{c}{$c_3$}  \\

  \hline

0.265 & $12-20$ &0.19 & $-0.23(1)$ & $12-20$& 0.15& $-0.18(7)$ & -5(7) & $9-20$ &0.32 & $-0.67(7)$ & 110(1)  & $-8429(451)$ \\
0.27  & $11-20$ &0.08 & $-0.19(1)$ & $12-20$& 0.71& $-0.32(7)$ & 14(6) & $9-20$ &0.96 & $-0.81(8)$ & 125(11) & $-7868(475)$ \\
0.275 & $13-20$ &0.22 & $-0.19(2)$ & $12-20$& 0.94& $-0.41(6)$ & 26(7) & $9-20$ &0.88 & $-0.81(8)$ & 118(11) & $-6551(479)$ \\
0.28  & $13-20$ &0.23 & $-0.18(2)$ & $11-20$& 0.33& $-0.32(5)$ & 17(4) & $8-20$ &0.41 & $-0.64(5)$ & 85(6)   & $-4333(207)$ \\
0.285 & $13-20$ &0.16 & $-0.18(2)$ & $11-20$& 0.37& $-0.35(5)$ & 21(4) & $8-20$ &0.93 & $-0.63(5)$ & 78(6)   & $-3620(209)$ \\
0.29  & $13-20$ &0.06 & $-0.18(2)$ & $10-20$& 0.08& $-0.30(3)$ & 18(3) & $7-20$ &0.43 & $-0.55(3)$ & 64(3)   & $-2673(78) $ \\
0.295 & $13-20$ &0.28 & $-0.17(2)$ & $10-20$& 0.25& $-0.30(3)$ & 18(3) & $7-20$ &0.83 & $-0.52(3)$ & 59(3)   & $-2294(75) $ \\
0.3   & $13-20$ &0.26 & $-0.16(2)$ & $10-20$& 0.11& $-0.31(3)$ & 19(3) & $7-20$ &0.84 & $-0.52(3)$ & 57(3)   & $-2036(77) $ \\
0.305 & $13-20$ &0.35 & $-0.16(2)$ & $10-20$& 0.14& $-0.31(3)$ & 21(3) & $7-20$ &0.60 & $-0.50(3)$ & 53(3)   & $-1765(73) $ \\
0.310 & $13-20$ &0.05 & $-0.15(2)$ & $10-20$& 0.09& $-0.31(3)$ & 22(3) & $7-20$ &0.46 & $-0.46(3)$ & 48(3)   & $-1514(77) $ \\
0.315 & $13-20$ &0.33 & $-0.14(2)$ & $9 -20$& 0.13& $-0.24(2)$ & 16(2) & $6-20$ &0.46 & $-0.39(2)$ & 40(2)   & $-1179(29) $ \\
0.32  & $13-20$ &0.21 & $-0.14(2)$ & $10-20$& 0.31& $-0.31(3)$ & 23(3) & $6-20$ &0.39 & $-0.39(2)$ & 39(2)   & $-1070(30) $ \\

  \hline\hline
 \end{tabular}
%}
\end{sidewaystable}

\begin{sidewaystable}
   \caption{ Results for the $111$ interface tension using of fits
according to
Eqs.~(\ref{fit_2}), (\ref{fit_3}), and (\ref{fit_5}), respectively.
}
  \label{tab:s111}
 \begin{tabular}{llllrlllll}
  \hline\hline
  &\multicolumn{3}{c}{fit ansatz (\ref{fit_2})}
  &\multicolumn{3}{c}{fit ansatz (\ref{fit_3})}
  &\multicolumn{3}{c}{fit ansatz (\ref{fit_5})}\\
  \multicolumn{1}{c}{$\beta$}
  &\multicolumn{1}{c}{fit range}
  &\multicolumn{1}{c}{$Q$}
  &\multicolumn{1}{c}{$\sigma_{111}$}
  &\multicolumn{1}{c}{fit range}
  &\multicolumn{1}{c}{$Q$}
  &\multicolumn{1}{c}{$\sigma_{111}$}
  &\multicolumn{1}{c}{fit range}
  &\multicolumn{1}{c}{$Q$}
  &\multicolumn{1}{c}{$\sigma_{111}$}\\
  \hline

  0.275 & $15-20$ &0.61 & 0.73746(8) & $14-20$& 0.05&0.73610(27) & $13-20$& 0.23&0.74114(79)\\
  0.28  & $14-20$ &0.20 & 0.80000(6) & $14-20$& 0.51&0.79990(8)  & $13-20$& 0.53&0.80230(79)\\
  0.285 & $14-20$ &0.93 & 0.86027(6) & $14-20$& 0.82&0.86030(27) & $12-20$& 0.09&0.86188(79)\\
  0.29  & $13-20$ &0.82 & 0.91865(5) & $13-20$& 0.70&0.91859(17) & $12-20$& 0.19&0.91951(81)\\
  0.295 & $13-20$ &0.65 & 0.97491(5) & $13-20$& 0.67&0.97506(17) & $12-20$& 0.79&0.97660(45)\\
  0.3   & $13-20$ &0.31 & 1.02927(5) & $13-20$& 0.75&1.02955(16) & $11-20$& 0.25&1.03140(27)\\
  0.305 & $12-20$ &0.07 & 1.08180(4) & $12-20$& 0.05&1.08190(12) & $11-20$& 0.45&1.08362(27)\\
  0.31  & $12-20$ &0.07 & 1.13242(4) & $12-20$& 0.39&1.13267(11) & $10-20$& 0.39&1.13429(17)\\
  0.315 & $12-20$ &0.06 & 1.18133(4) & $12-20$& 0.86&1.18168(12) & $10-20$& 0.77&1.18275(18)\\
  0.32  & $14-20$ &0.27 & 1.22872(6) & $12-20$& 0.46&1.22900(12) & $10-20$& 0.69&1.22971(18)\\
  \hline\hline
 \end{tabular}
\end{sidewaystable}

\begin{sidewaystable}
  \caption{Results for the parameters $c_i$ for the $111$ interface tension 
from fits according to Eqs.~(\ref{fit_2}), (\ref{fit_3}), and (\ref{fit_5}), respectively.
}
  \label{tab:s111ci}
 \begin{tabular}{lllllllllllll}

  \hline\hline

  &\multicolumn{3}{c}{fit ansatz (\ref{fit_2})}
  &\multicolumn{4}{c}{fit ansatz (\ref{fit_3})}
  &\multicolumn{5}{c}{fit ansatz (\ref{fit_5})}\\

   \multicolumn{1}{c}{$\beta$}
  &\multicolumn{1}{c}{fit range}
  &\multicolumn{1}{c}{$Q$}
  &\multicolumn{1}{c}{$c_1$}
  &\multicolumn{1}{c}{fit range}
  &\multicolumn{1}{c}{$Q$}
  &\multicolumn{1}{c}{$c_1$}
  &\multicolumn{1}{c}{$c_2$}
  &\multicolumn{1}{c}{fit range}
  &\multicolumn{1}{c}{$Q$}
  &\multicolumn{1}{c}{$c_1$}
  &\multicolumn{1}{c}{$c_2$}
  &\multicolumn{1}{c}{$c_3$}  \\

  \hline
  0.275 & $15-20$ &0.61 & $-0.44(3)$ & $14-20$& 0.05& $0.38(15) $ & $-120(19)$ & $13-20$& 0.23& $-3.8(6)$ & 1024(147) & $-99709(11487)$ \\
  0.28  & $14-20$ &0.20 & $-0.42(2)$ & $14-20$& 0.51& $-0.15(15)$ & $-34(19) $ & $13-20$& 0.53& $-2.5(6)$ & 604(146)  & $-56031(11363)$ \\
  0.285 & $14-20$ &0.93 & $-0.34(1)$ & $14-20$& 0.82& $-0.36(14)$ & $ 2(19)  $ & $12-20$& 0.09& $-2.9(3)$ & 678(71)   & $-57114(4935) $ \\
  0.29  & $13-20$ &0.82 & $-0.32(2)$ & $13-20$& 0.70& $-0.36(6) $ & $ -29(6) $ & $12-20$& 0.19& $-2.2(4)$ & 465(71)   & $-37386(4952) $ \\
  0.295 & $13-20$ &0.65 & $-0.27(2)$ & $13-20$& 0.67& $-0.37(8) $ & $ 17(9)  $ & $12-20$& 0.79& $-1.5(4)$ & 307(70)   & $-23693(4885) $ \\
  0.3   & $13-20$ &0.31 & $-0.23(1)$ & $13-20$& 0.75& $-0.37(8) $ & $ 17(10) $ & $11-20$& 0.25& $-1.8(2)$ & 360(35)   & $-26311(2109) $ \\
  0.305 & $12-20$ &0.07 & $-0.21(1)$ & $12-20$& 0.05& $-0.25(5) $ & $ 5(5)   $ & $11-20$& 0.45& $-1.5(2)$ & 289(34)   & $-20160(2070) $ \\
  0.31  & $12-20$ &0.07 & $-0.17(1)$ & $12-20$& 0.39& $-0.29(5) $ & $ 12(5)  $ & $10-20$& 0.39& $-1.4(1)$ & 270(17)   & $-17989(868)  $ \\
  0.315 & $12-20$ &0.06 & $-0.15(1)$ & $12-20$& 0.86& $-0.31(5) $ & $ 17(6)  $ & $10-20$& 0.77& $-1.1(1)$ & 191(17)   & $-12349(892)  $ \\
  0.32  & $14-20$ &0.27 & $-0.13(1)$ & $12-20$& 0.46& $-0.34(5) $ & $ 22(6)  $ & $10-20$& 0.69& $-0.9(1)$ & 143(18)   & $-8827(930)   $ \\
  \hline\hline
 \end{tabular}
\end{sidewaystable}


\newpage
\begin{thebibliography}{MM}

\bibitem{droplet}
J. E. Avron, H. van Beijeren, L. S. Schulman, and R. K. P.  Zia,
%  Roughening transition, surface tension and equilibrium droplet shapes in a two-dimensional Ising system
J. Phys. A: Math. Gen. 15, L81 (1982).

\bibitem{rough}
M. Hasenbusch and K. Pinn, J. Phys. A 30, 63 (1997).

\bibitem{wulff}
G. Wulff, Z. Kristallogr. Mineral. 34, 449 (1901).

\bibitem{3d_s100_binder}
K. Binder, Phys. Rev. A 25, 1699 (1982).

\bibitem{3d_s100_berg}
B. A. Berg, U. Hansmann, and T. Neuhaus, Z. Phys. B 90, 229 (1993).

\bibitem{hasenbusch94}
M. Hasenbusch and K. Pinn, Physica A 203, 189 (1994).

\bibitem{3d_s100_caselle94}
M. Caselle, R. Fiore, F. Gliozzi, M. Hasenbusch, K. Pinn, and S. Vinti, 
Nucl. Phys. B 432, 590 (1994).

\bibitem{hasenbusch97}
M. Hasenbusch and K. Pinn, Physica A 245, 366 (1997).

\bibitem{3d_s100_chatelain}
C. Chatelain, J. Stat. Mech. P04011 (2007).


\bibitem{3d_s100_caselle06}
M. Caselle, M. Hasenbusch,  and M. Panero,
JHEP 03, 084 (2006);
JHEP 09, 117 (2007).


\bibitem{3d_s110}
K. K. Mon, S. Wansleben, D. P. Landau, and K. Binder,
%Anisotropic Surface Tension, Step Free Energy and Interfacial Roughening in the 3D Ising Model
Phys. Rev. Lett. 60, 708 (1988);
% Monte Carlo studies of anisotropic surface tension and interfacial roughening in the 3D Ising Model
Phys. Rev. B 39, 7089 (1989).

\bibitem{bloete99}
H. W. J. Bl\"ote, L. N. Shchur, and A. L. Talapov, 
Int. J. Mod. Phys. C 10, 1137 (1999).

\bibitem{ebanwj}
E. Bittner, A. Nu{\ss}baumer, and W. Janke, in preparation.

\bibitem{wj_nic}
W.~Janke, {\it Histograms and all that\/},
in: {\it Computer Simulations of Surfaces and Interfaces\/},
NATO Science Series, II.\ Mathematics, Physics and Chemistry -- Vol.~{\bf 114},
%Proceedings of the NATO Advanced Study Institute, 
%Albena, Bulgaria, %September 2002, 
edited by B.~D\"unweg, D. P.~Landau, and A. I.~Milchev
(Kluwer, Dordrecht, 2003); pp.~137--157.

\bibitem{berg96}
B. A. Berg, J. Stat. Phys. 82, 323 (1996).

\bibitem{binder81}
K. Binder, Z. Phys. B 43, 119 (1981). 

\bibitem{brezin85}
E. Brezin and J. Zinn-Justin,
Nucl. Phys. B 257, 867 (1985). 

\bibitem{gelfand90}
M. P. Gelfand and M. E. Fisher, 
Physica A 166, 1 (1990).

\bibitem{morris91}
J. J. Morris,
J. Stat. Phys. 69, 539 (1991).

\bibitem{leung:90}
K. Leung and R. K. P. Zia, 
J. Phys. A: Math. Gen. 23, 4593 (1990).

\bibitem{neuhaus.hager}
T. Neuhaus and J. S. Hager,
J. Stat. Phys. 113, 47 (2003).

\bibitem{binder03}
K. Binder, Physica A 319, 99 (2003).

\bibitem{nussbaum}
A. Nu{\ss}baumer, E. Bittner, and W. Janke, 
Phys. Rev. E 77, 041109 (2008).

\bibitem{onsager}
L. Onsager, Phys. Rev. 65, 117 (1944).

\bibitem{baxter_book}
R. Baxter, {\it Exactly Solved Models in Statistical Mechanics\/}
(Academic Press, London, 1982).




\end{thebibliography}
\end{document}